\documentclass[journal]{IEEEtran}
\usepackage{graphics} 
\usepackage{epsfig} 
\usepackage{times} 
\usepackage{soul}
\usepackage{amsmath} 
\usepackage{amssymb}  
\usepackage{subfigure}
\usepackage{url}
\usepackage{arydshln}
\newtheorem{theorem}{Theorem}

\newtheorem{algorithm}[theorem]{Algorithm}

\usepackage{color}
\usepackage{algorithm}

\newtheorem{definition}[theorem]{Definition}

{ 

}

\newcommand{\bi}{\begin{itemize}}
\newcommand{\ei}{\end{itemize}}
\newcommand{\bd}{\begin{displaymath}}
\newcommand{\ed}{\end{displaymath}}
\newcommand{\be}{\begin{eqnarray*}}
\newcommand{\ee}{\end{eqnarray*}}

\ifCLASSINFOpdf

\else

\fi

\begin{document}

\title{On Information Transfer Based Characterization of Power System Stability}
\author{Subhrajit Sinha,
        Pranav Sharma,
        Umesh Vaidya
        and Venkataramana Ajjarapu 
}


\maketitle

\begin{abstract}

In this paper, we present a novel approach to identify the generators and states responsible for the small-signal stability of power networks. To this end, the newly developed notion of information transfer between the states of a dynamical system is used. In particular, using the concept of information transfer, which characterizes influence between the various states and a linear combination of states of a dynamical system, we identify the generators and states which are responsible for causing instability of the power network. While characterizing influence from state to state, information transfer can also describe influence from state to modes thereby generalizing the well-known notion of participation factor while at the same time overcoming some of the limitations of the participation factor. The developed framework is applied to study the three bus system identifying various cause of instabilities in the system. The simulation study is extended to IEEE 39 bus system.

\end{abstract}

\IEEEpeerreviewmaketitle

\section{Introduction}

The electric power grid is one of the largest network systems with thousands of components having complicated dynamics and topology. As such, control and maintaining reliable operation of a power network is challenging and difficult and one of the main challenges is the maintenance of stability of power networks. For a large network, it is usually very difficult to identify the sources of instability and hence an \emph{a priori} knowledge of the sources of instability would allow one to take precautionary measures before a power network becomes unstable.

Power system stability is the ability of the system to reach an operating equilibrium after disturbance from an initial operating condition where most system variables remain bounded in their practical limits \cite{kundur_stability_classification}. Inherently, instability is a single phenomenon, however for the ease of understanding, instabilities are classified based on time scale, magnitude and parameters causing them. In order to understand the behavior of a complex power network, we consider the linear behavior of the system in close affinity of an operating equilibrium. This study of power system is categorized as \textit{small signal analysis} \cite{kundur_stability_classification}. Small signal analysis capture system behavior and interaction of dynamic components without analyzing complex non-linear system models. In  \cite{perez1982selective,verghese1982selective} the authors defined the concept of \textit{participation factor} to analyze these linearized dynamic models. Participation factor is based on eigenvalue analysis and defines the influence of any single state on any system mode. This method is thus far prevalent in power system studies for model reduction, optimal control placement and identifying resonance interaction \cite{abed2000participation}.
However, in \cite{hashlamoun2009new, abed2009modal} authors identified anomalies in participation factor definition and showed how the existing definition fails in even two dimensional linear systems. The authors further provided a new definition for participation factor which accounts for uncertainties in system states, which increases the computation involved in participation factor analysis. The other drawback associated with existing framework for modal analysis is that it only quantifies participation of an individual state into each mode. For a realistic power network with a huge number of dynamical states, it becomes computationally challenging to evaluate system behavior at a macro level \cite{van1994interaction}. Thus a method is needed, where influence of physical components and subsystems connected in a power system can be considered collectively.

On the other hand, causality analysis has been a topic of research from the days of Aristotle, but there has not been a universal definition of causality. Clive Granger provided his definition of causality \cite{granger_economics,granger_causality} and since then it has become one of the most popular tests for causality detection. Again, Massey and Kramer \cite{IT_massey_directed,IT_kramer_directedit} generalized Shannon's information theory \cite{Shannon_1948} to incorporate a sense of direction and thus defined directed information. Apart from these, there are other measures of causality like transfer entropy \cite{Schreiber} and Liang-Kleeman information \cite{liang_kleeman_prl}. However, in \cite{IT_causality_cdc}, it was shown that in dynamical system setting, all these measures fail to capture the true notion of causality and in \cite{IT_causality_cdc,sinha2017information}, the authors provided a new definition of causality and showed how the new definition captures the true causal structure in a dynamical system. The idea of causality and influence is based on the concept of information transfer in a dynamical system and the identified causal structure and influence characterization can be used for stability analysis \cite{IT_influence_acc,sinha_cdc_2017_power}. The results in \cite{IT_influence_acc,sinha_cdc_2017_power} are based on eigenvalue and eigenvector decomposition of the linearized model of the system and as such, as the size of the power network grows, it becomes computationally challenging.

In this paper, we provide a novel way for stability characterization for power network using the notion of information transfer between dynamic states. Information transfer quantifies the influence between any state(s) and any other state(s), that is, how the evolution of a state(s) affect the evolution of any other state(s). This causality characterization act as a measure to  identify \textit{'when'} system can undergo instability and \textit{'who'} is responsible for it. In power system case, information transfer identifies operating condition and dynamic states (and generators) which are responsible for instability. The key advantage of this method is that it does not require the computation of eigenvalues and eigenvectors and is thus computationally less expensive than the previous approaches. Furthermore, information transfer quantifies the interaction of subspaces (combination of states) in a dynamical system, which helps in understanding the system behavior along the lines of physical properties of system and phenomenon to be observed. In particular, information transfer measure allows system states to be clubbed together so that system properties can be analyzed and inferred at a macro level, without focusing on an individual state. This reduces the computational complexity of analysis for large scale systems.

The structure of this paper is as follows. In section \ref{section_IT}, we review the concept of information transfer and provide a physical interpretation for stability analysis and participation from information transfer standpoint. In section \ref{section_IT_PF} we discuss the advantages of information transfer over existing method of eigenvalue analysis and participation factor. In section \ref{section_3_bus} we take a tutorial example of three bus system to explain the notion of information transfer for power system analysis for identifying states responsible for instability. Further, in section \ref{section_39_bus} we analyze the IEEE 39 bus system for identifying instabilities and responsible states (and generators). Finally we conclude the paper in section \ref{section_conclusion}, while highlighting the future implications of this work.

\section{Information Transfer Based Participation and Stability Analysis}\label{section_IT}
In this section, we briefly describe the concept of information transfer in a dynamical system \cite{IT_causality_cdc}, \cite{sinha2017information}. 

Consider the discrete time dynamical system
\begin{eqnarray}\label{system}
z(t+1) = S(z(t)) + \xi(t)
\end{eqnarray}
where $z = \begin{pmatrix}
x^\top & y^\top
\end{pmatrix}^\top \in \mathbb{R}^N$, $S:\mathbb{R}^N\to \mathbb{R}^N$ is assumed to be at least continuous and $\xi(t)$ is independent and identically distributed additive noise, which comes from the distribution $g$. 

Information transfer from state (subspace) $x$ to state (subspace) $y$ gives a measure of how the evolution of $x$ dynamics affect (influence) the evolution of $y$ dynamics. In particular, we quantify this influence in terms of the entropy transferred from the $x$ dynamics to the $y$ dynamics, as the system (\ref{system}) evolves in time. Note that, by entropy we mean the Shannon entropy, which is defined for a probability distribution as follows :

\begin{definition}[Shannon Entropy]
Given a probability distribution $\rho(z)$, defined over the sample space $\Omega$, the entropy of the distribution $(H)$ is defined as 
\begin{eqnarray}\label{entropy_def}
H = -\int_{\Omega}\rho(z)\log\rho(z)dz
\end{eqnarray}
\end{definition}

The entropy of a distribution is the measure of the information content of the distribution. Suppose there are two agents (states in the case of a dynamical system) which are interacting with each other. Each has its own entropy (information) and they \emph{transfer} a part of their own information to the other agent via the interaction. We use this intuition to define the information transfer. In particular, information transfer from $x$ to $y$ is the amount of information (entropy) of $x$ that is being transferred to $y$, the system (\ref{system}) evolves in time. With this, we define the information transfer as follows.

\begin{definition}[Information Transfer]
The information transfer from a state $x$ to state $y$ $(T_{x\to y})$, as the dynamical system $z(t+1) = S(z(t)) + \xi(t)$
evolves from time step $t$ to time step $t+1$ is defined as
{\small
\begin{eqnarray}\label{IT_def}
T_{x\to y} = H(y(t+1)|y(t))-H_{\not{x}}(y(t+1)|y(t))
\end{eqnarray}}
where $z = \begin{pmatrix}
x & y
\end{pmatrix}^\top$, $H(y(t+1)|y(t))$ is the entropy of $y$ at time $t+1$ conditioned on $y(t)$ and $H_{\not{x}}(y(t+1)|y(t))$ is the conditional entropy of $y(t+1)$, conditioned on $y(t)$, when $x$ is absent from the dynamics.
\end{definition}
The intuition behind the definition of information transfer is the fact that the total entropy of $y$ is the entropy of $y$ when $x$ is absent from the dynamics plus the entropy transferred from $x$ to $y$. Hence, the information transfer from $x$ to $y$ gives the amount of entropy flowing from $x$ to $y$ and thus quantifies how much the $x$ dynamics affects the $y$ dynamics. With this, we define influence in a dynamical system as follows,
\begin{definition}[Influence]
We say a state (or subspace) $x$ influences a state (or subspace) $y$ if and only if the information transfer from $x$ to $y$ is non-zero.
\end{definition}
Larger the absolute value of the $T_{x\to y}$, more is the effect of $x$ dynamics on $y$ and hence more is the influence of $x$ on $y$. 
For general nonlinear systems it is often not possible to find close-form expressions for evolution of entropy, and hence the information transfer can be calculated using numerical techniques. However, for linear systems of the form
\begin{eqnarray}
z(t+1)=Az(t)+\sigma \xi(t)\label{lti}
\end{eqnarray}
where $z(t)\in \mathbb{R}^N$, $\sigma > 0$ is a constant and $\xi(t)$ is vector valued Gaussian random variable with zero mean and unit variance, the information transfer from any state $z_i$ to $z_j$ can be expressed as a closed form expression. In particular, we have the following theorem \cite{IT_causality_cdc,sinha2017information}.
\begin{theorem}\label{IT_theorem}
Consider the linear dynamical system (\ref{lti}). We have the following expression for information transfer between various subspaces 

\begin{eqnarray}
[T_{x_1\to y}]_t^{t+1}=\frac{1}{2}\log \frac{|A_{yx}\Sigma^s_y(t)A_{yx}^\top +\sigma^2 I |}{|A_{yx_2}(\Sigma_y^{s})_{yx_2}(t)A_{yx_2}^\top+\sigma^2 I|}\label{transferx1y}
\end{eqnarray}

where 
\begin{eqnarray}
A = \begin{pmatrix}A_x&A_{xy}\\ A_{yx}&A_{y}\end{pmatrix}=\begin{pmatrix}A_{x_1}&A_{x_1x_2}& A_{x_1 y}\\A_{x_2x_1}&A_{x_2}& A_{x_2 y}\\ A_{y x_1}&A_{y x_2}& A_{y}\end{pmatrix},\label{splittingA}
\end{eqnarray}
\begin{eqnarray}
\Sigma=\begin{pmatrix}\Sigma_x&\Sigma_{xy}\\\Sigma_{xy}^\top& \Sigma_y\end{pmatrix}=\begin{pmatrix} \Sigma_{x_1}&\Sigma_{x_1x_2}&\Sigma_{x_1 y}\\\Sigma_{x_1x_2}^\top&\Sigma_{x_2}&\Sigma_{x_2 y}\\\Sigma_{x_1y}^\top&\Sigma_{x_2y}^\top&\Sigma_{y}\end{pmatrix},
\label{sigma_dec}
\end{eqnarray}

$\Sigma^s_y(t)=\Sigma_x(t)-\Sigma_{xy}(t)\Sigma_y(t)^{-1}\Sigma_{xy}(t)^\top$ is the Schur complement of $\Sigma_{y}(t)$ in the matrix $\Sigma(t)$, $|\cdot|$ is the determinant and $ (\Sigma_y^s)_{yx_2}$ is the Schur complement of $\Sigma_{y}$ in the matrix 
\[\begin{pmatrix}\Sigma_{x_2}&\Sigma_{x_2y}\\\Sigma_{x_2 y}^\top&\Sigma_y\end{pmatrix}.\]

Setting, $z_i=x_1$, $z_j = y$ and $\{z_1,\cdots ,z_N\}\setminus \{z_i,z_j\}=x_2$, one obtains the information transfer from any state $z_i$ to any state $z_j$.

\end{theorem}

The information transfer thus defined, can be extended to define information transfer between the various signals in a control dynamical system, namely information transfer from input to state, input to output and state to output. For details see \cite{IT_causality_cdc}.

\begin{figure}[htp!]
\centering
\includegraphics[scale=.35]{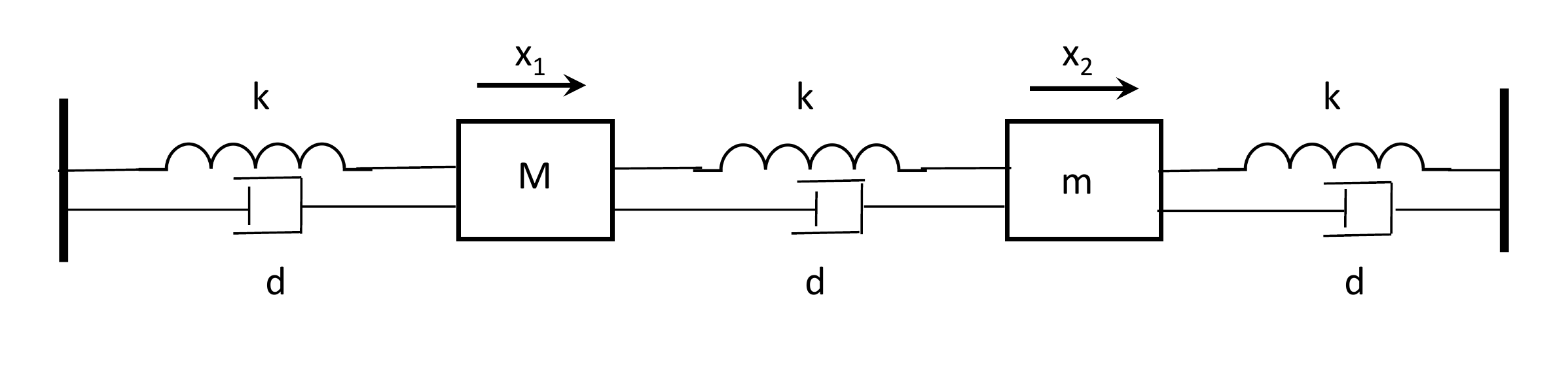}
\caption{Mass-spring-damper system}\label{mass_spring_fig}
\end{figure}

To convey the meaning of information transfer, we consider a mass-spring-damper system, as shown in Fig. \ref{mass_spring_fig}.

The equations of motion for the mass-spring system are 
\begin{eqnarray}
M\ddot{x}_1 + 2d\dot{x}_1 - d\dot{x}_2 + 2kx_1 - kx_2 = 0\\
m\ddot{x}_2 + 2d\dot{x}_2 - d\dot{x}_1 + 2kx_2 - kx_1 = 0
\end{eqnarray}
where $M,m$ are the masses, $d$ is the damping coefficient and $k$ is the spring constant. We assume that the damping coefficients of the dampers are equal and so are the spring constants of the springs.
In state space form, the system can be represented as
{\small
\begin{eqnarray*}
\begin{pmatrix}
\dot{z}_1\\
\dot{z}_2\\
\dot{z}_3\\
\dot{z}_4
\end{pmatrix} = \begin{pmatrix}
0 & 1 & 0 & 0\\
-2k/M & -2d/M & k/M & d/M\\
0 & 0 & 0 & 1\\
k/m & d/m & -2k/m & -2d/m
\end{pmatrix}\begin{pmatrix}
{z}_1\\
{z}_2\\
{z}_3\\
{z}_4
\end{pmatrix}
\end{eqnarray*}
}

where $z_1 = x_1$, $z_2 = \dot{x}_1$, $z_3 = x_2$ and $z_4=\dot{x}_2$.
For simulation purposes, we choose $M=10$, $m = 3$, $k=20$ and $d=1.5$ with appropriate units.  Since $M>m$, a perturbation (perturbed so that it has some non-zero initial position) in $M$ will result in larger oscillations in the masses, compared to the case when $m$ is perturbed by the same amount. Hence, we can conclude that $M$ has a large influence on $m$, whereas, $m$ has a much smaller influence on $M$. 

\begin{figure}[htp!]
\centering
\subfigure[]{\includegraphics[scale=.22]{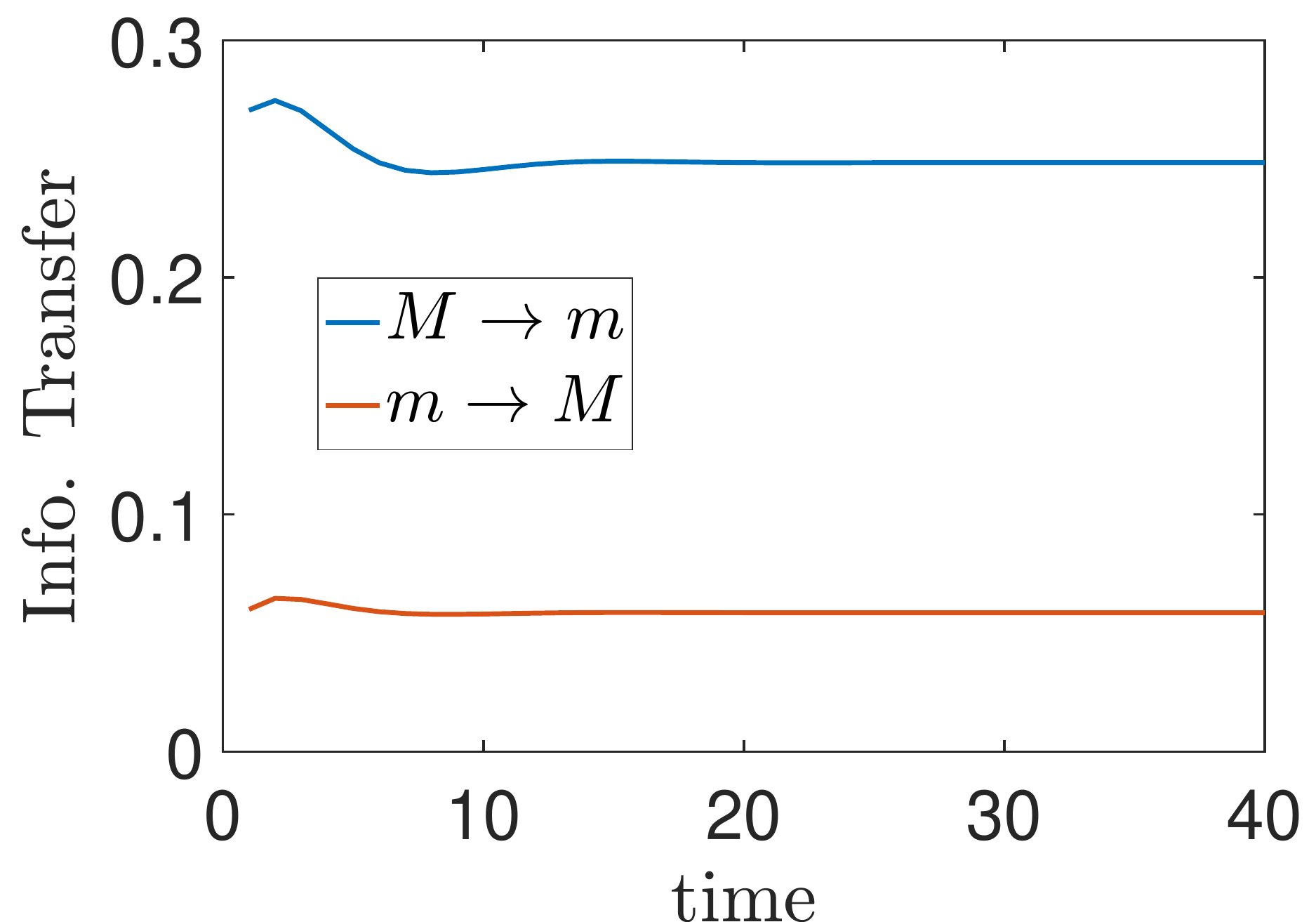}}
\subfigure[]{\includegraphics[scale=.35]{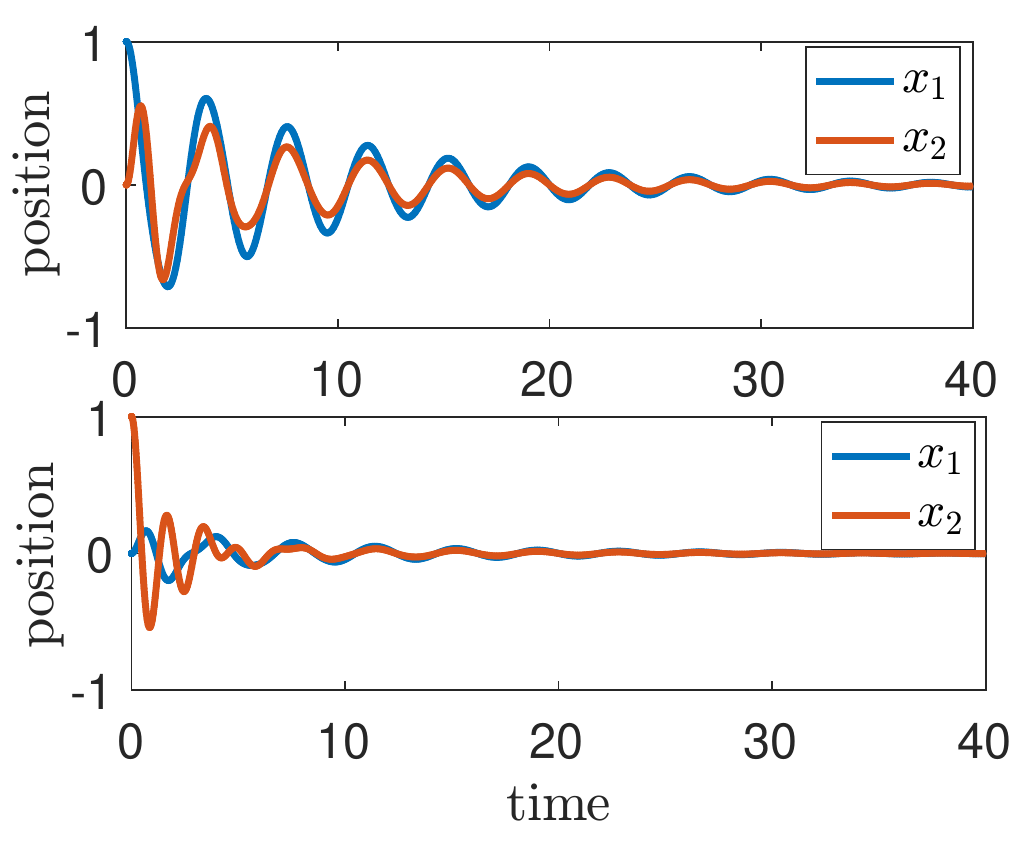}}
\caption{(a) Information transfer between the two masses. (b) The top figure shows the position of the two masses when the bigger mass is at an initial position of 1 unit and the lower figure shows the position of the two masses when the smaller mass is at an initial position of 1 unit.}\label{IT_mass_spring}
\end{figure}

Information transfer between the two masses is shown in Fig. \ref{IT_mass_spring}(a). The information transfer plotted are from $(z_1,z_2)$ subspace to $(z_3,z_4)$ subspace and vice versa. As can be seen from the figure, the larger mass has a greater influence on the smaller mass, as compared to the influence of smaller mass on the larger mass. This is also verified from the time-domain plots in Fig. \ref{IT_mass_spring}(b). In both the plots, we plot the position of the two masses. In the top figure, the larger mass was initialized at 1 unit (that is, its initial position was set to 1 unit), and the system was allowed to evolve from this initial condition. In the bottom figure, the initial position of the smaller mass was initialized at 1 unit and the system was allowed to evolve from this initial condition. From Fig. \ref{IT_mass_spring}(b), it can be observed that when the larger mass initialized at 1 unit, oscillations sustain for a longer time as compared to the case when the smaller mass is initialized at 1 unit. In fact, when $M$ is initialized to the non-zero initial condition, the variance of position of $m$ is $0.0313$, whereas when $m$ is initialized to the same non-zero initial condition the variance of position of $M$ is $0.0030$. Hence, time-domain simulations confirm the conclusions obtained using information transfer measure.

\subsection{Information Transfer and Stability}
\begin{figure}[htp!]
\centering
\includegraphics[scale=.325]{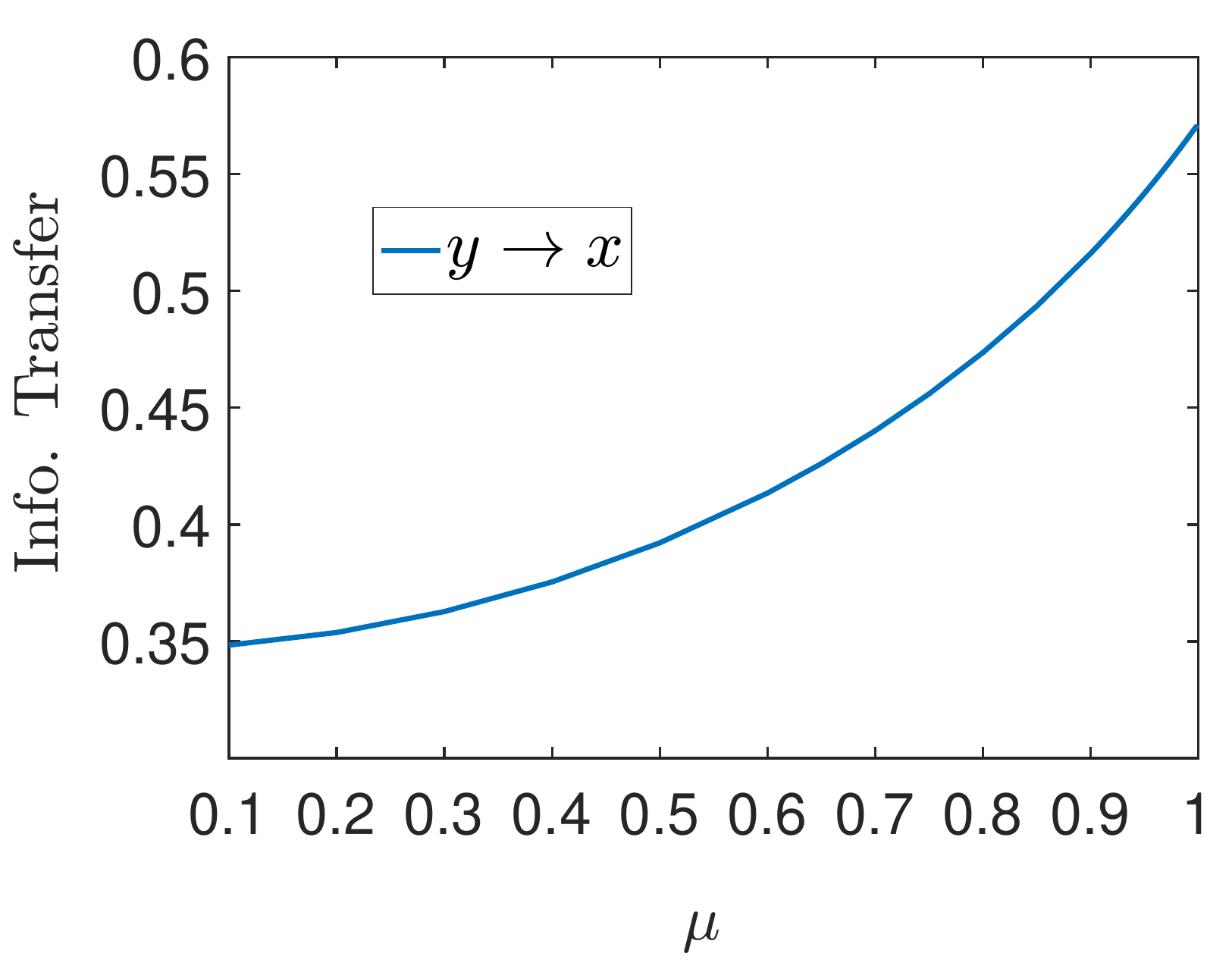}
\caption{Information transfer increases rapidly as the system approaches instability.}\label{IT_stability}
\end{figure}

As discussed earlier, information transfer can be used to measure the influence of one state variable on another state. However, the state to state information transfer can also be used as an indicator of the instability of a system. In particular, information transfer can be used to identify the state which makes a system become more vulnerable to perturbations. For example, consider a linear system
\begin{eqnarray}\label{sys_IT_stability}
\begin{aligned}
& x(t+1) = 0.7x(t) + y(t) + \xi_x(t)\\
& y(t+1) = \mu y(t) + \xi_y(t)
\end{aligned}
\end{eqnarray}
where $\mu\in [0.1,0.99]$ and $\xi_x(t)$ and $\xi_y(t)$ are independent and identical Gaussian noises of unit variance. The eigenvalues of the system are $(0.7,\mu)$ and hence as $\mu$ approaches 1, the system approaches instability. The instability occurs due to $y$ dynamics and as $\mu$ increases, the entropy of $y$ increases rapidly. Hence, the information transfer from $y$ to $x$ also increases rapidly. This is shown in Fig. \ref{IT_stability}. 
Conversely, if the information transfer from some state (subspace) to any other state (subspace) increases rapidly, it can be concluded that the system is approaching instability. The point to be noted is that the rapid increase in the information transfer happens when the system is still operating in the stable zone. Hence, can use information transfer to predict the onset of instability and hence one can take preventive measures before the system becomes unstable. 


\section{Information Transfer and Participation Factor}\label{section_IT_PF}
Participation matrix $P$ \cite{participation_part1} for a linear system
\begin{eqnarray}\label{sys_lin}
\dot{x} = Ax
\end{eqnarray}
is defined as 
\begin{eqnarray}
P = [p]_{ki} = {\bf u}_k^i {\bf w}_k^i
\end{eqnarray}
where ${\bf u}_k^i$  (${\bf w}_k^i$) is the $k^{th}$ component of the $i^{th}$ left (right) eigenvector corresponding to eigenvalue $\lambda_i$ of the power system $A$ matrix (Eq. \ref{sys_lin}). These eigenvectors are assumed to be normalized i.e., \[{\bf u}_i^\top {\bf w}_j=1,\;\; {\rm if}\;\; i=j;\;0\;\;{\rm otherwise}.\] 
Participation factor $p_{ki}$ measures the relative participation of $k^{th}$ state variable in the $i^{th}$ mode. Hence, larger the participation factor of a state to a particular mode, larger the contribution of that state to that mode. However, as pointed in \cite{abed2000participation,abed2009modal,abed_participation} participation factor fails to capture the contribution of states in some linear systems as illustrated in the following example highlighted in \cite{abed2009modal}. 

\begin{eqnarray}\label{2dim}
 \dot{x} = \begin{pmatrix}
-0.2231 & 3.4657\\
0 &  -0.9163
\end{pmatrix}x \end{eqnarray}
where $x\in\mathbb{R}^{2}$.

The eigenvalues of the system matrix are $\lambda_1 = -0.2231$ and $\lambda_2 = -0.9163$ and the participation matrix is
\begin{equation} 
P = 
\begin{pmatrix}
1 & 0\\
0 & 1
\end{pmatrix}
\end{equation}
Hence participation factor predicts that $x_1$ does not contribute in the mode $\lambda_2$ and eigenvector corresponding to $\lambda_2$ and $x_2$ does not contribute in mode $\lambda_1$ and the corresponding eigenvector. 
However,
\begin{eqnarray*}\label{2_state_evol}\nonumber
x_1(t) &=& [0.1961x_1(0) + 0.9806x_2(0)]e^{\lambda_1t} \\
&& - 0.9806e^{\lambda_2t}x_2(0)
\end{eqnarray*}
and so $x_2$ does \textit{participate} in $\lambda_1$. 
Information transfer, on the other hand illustrates the fact that $\lambda_1$ is being influenced by $x_2$. For this, we discretize the continuous time system, with $\delta t = 1$ and look at the information transfer from $x_2$ to the eigenvector corresponding to $\lambda_1$. For details on computation of information transfer from state to output see \cite{IT_influence_acc,sinha_IT_CDC2016,sinha_cdc_2017_power}.

\begin{figure}[htp!]
\centering
\includegraphics[scale=.5]{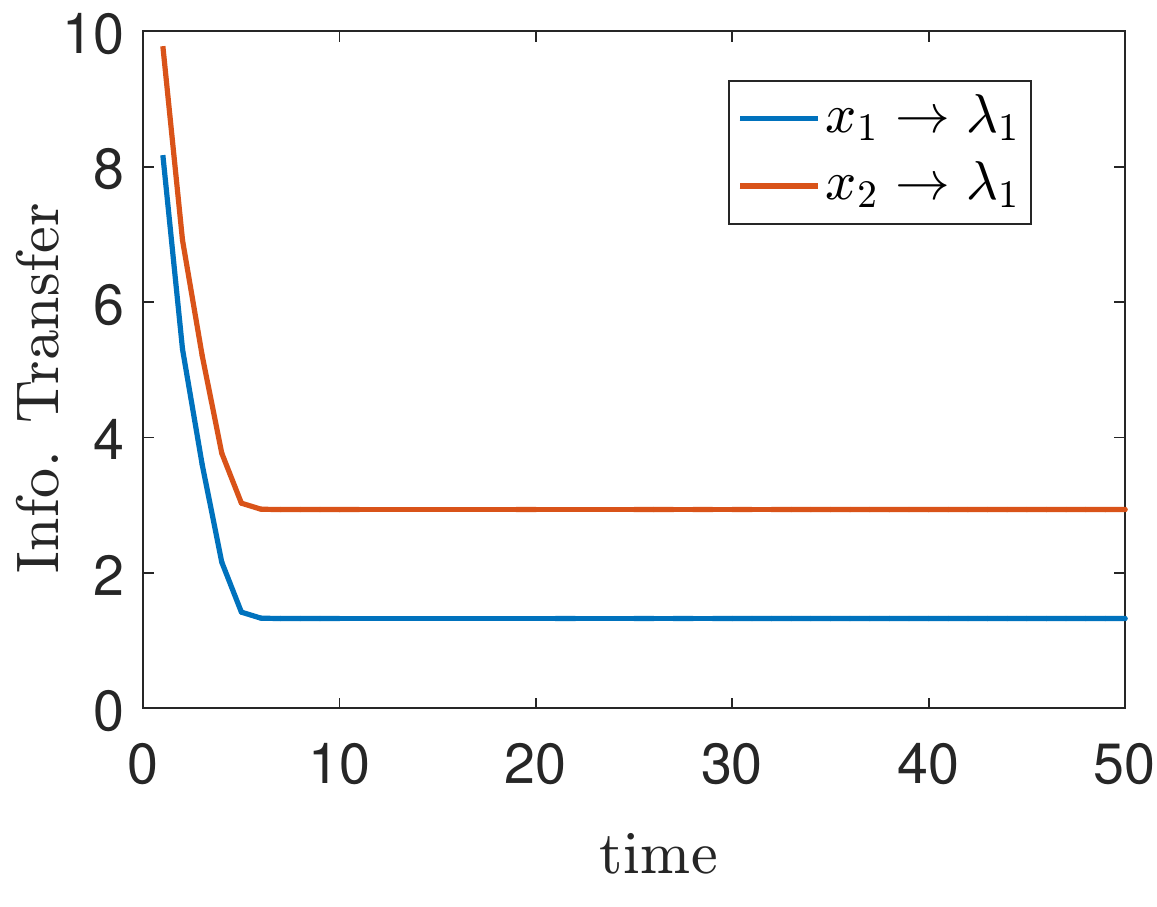}
\caption{Information Transfer from system states to $\lambda_1$}\label{2d_mode}
\end{figure}

Fig. \ref{2d_mode} shows the information transfer from the states to the mode $\lambda_1$ and it can be seen that the information flow $T_{x_2\to \lambda_1}$ is non-zero. Hence information transfer captures the influence of $x_2$ on mode $\lambda_1$.

\subsection{Subspace interaction in dynamic system}
Information transfer $T_{x \rightarrow  y}$, as defined in (\ref{IT_def}), is general enough in the sense that it not only quantifies the influence of one state on another but also from a combination of states (subspace) to any other combination of states (subspace). This definition has a great advantage in power system studies as it reduces the computational burden by utilizing the physical properties of the system and phenomenon. For example, in order to identify areas (generators) participating in an inter-area oscillatory mode using existing modal analysis method, one needs to compute all the eigenvalues and eigenvectors followed by participation factor of all the dynamic states of the system. We then select key states and thereby identify the generators and areas oscillating with respect to each other. However, as inter-area oscillations are properties of the bulk system, we can create subspaces by clustering all dynamic states in an area (or in a generator) into single clusters and then compute information transfer between these subspaces and inter-area modes. Information transfer directly identifies the key areas (generators) participating in a particular oscillatory mode. Thus, information transfer not only reduces the computation burden but also reveals the physical properties of the system in a more organic way. 
{\small
\begin{table*}[h!]
\centering
\caption{Computation Cost for methods of Modal Analysis}
\label{table_compute_cost}
\begin{tabular}{|c|c|c|c|c|}
\hline
Eigenpair computation method\textbackslash  &  Computation Cost & Description  \\
Information transfer & (for each iteration) & \\ \hline \hline
    QR Algorithm & $\mathbb{O}(n^3)$ & Factorize $A ( = QR)$ into an orthogonal and triangular matrix and\\ 
    & & apply the next iteration to RQ.  \\
\hline
    
     Divide and Conquer Algorithm & $\mathbb{O}(n^3)$ & Divide the matrix into subspaces of diagonalized matrices and then recombine.\\ \hline
    
Information Transfer& $\mathbb{O}(\bar{n}^2)$ & where $\bar{n} \in [log(n), n]$ depends upon the physical property of the system\\
& &or phenomena under observation.\\
     \hline
\end{tabular}
\end{table*}}

\subsection{Computational Cost}

The known method of modal analysis applied to small signal stability analysis involves computation of eigenvalues and eigenvectors. For a given linear system as given in equation (\ref{lti}), where $z(t) \in \mathbb{R}^n$, eigenvalue computation using known methods of Arnoldi iteration, QR algorithm, divide and conquer method and Jacobi eigenvalue algorithm is fairly complex. Eigenvalue computation involves an iterative method to obtain an equivalent diagonal or tridiagonal matrix for the given $A$ matrix. Further, the computation of eigenpair (eigenvalue and eigenvector) increases the computation complexity. As shown in Table \ref{table_compute_cost}, the given computation cost of eigenpair computation is for each iteration and the number of iterations depends greatly on the size and structure of matrix \cite{press2007numerical,li1994laguerre,golub2001eigenvalue,crow2015computational}. On the other hand information transfer computation is proportional to the order of $(\bar{n})^2$, where $(\bar n)$ depends upon the physical nature of the dynamic model and the phenomena under observation. As illustrated in the next section, for 39 bus system, we can reduce the number of dynamic subspaces, to the number of generators, in order to identify critical generators and dynamic states in the system. Details are illustrated in Table \ref{table_compute_cost}.

\color{black}
\section{Stability Characterization of 3-bus System}\label{section_3_bus}

\begin{figure}[htp!]
\centering
\subfigure[]{\includegraphics[scale=.45]{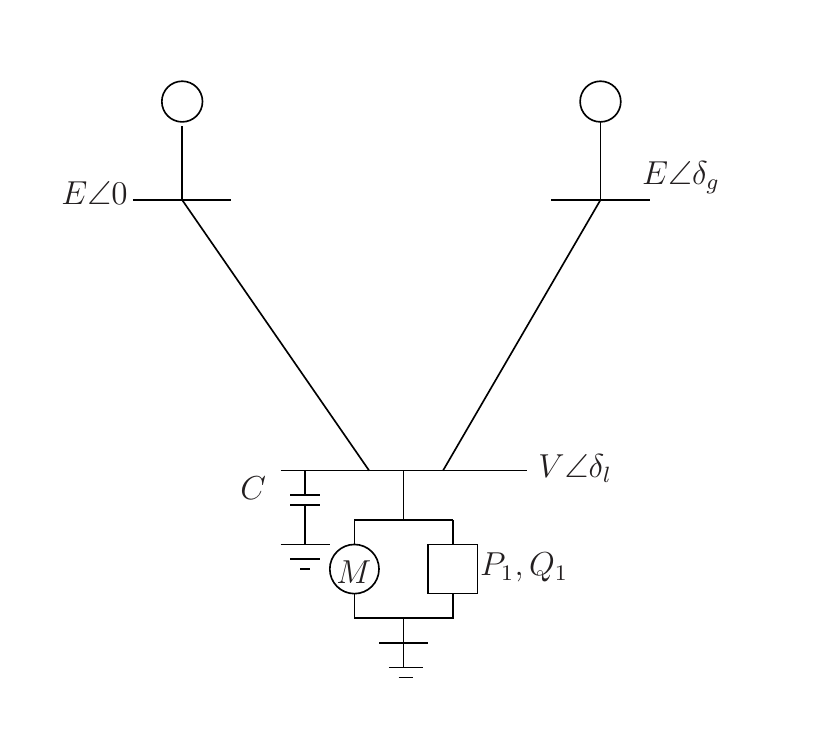}}
\subfigure[]{\includegraphics[scale=.24]{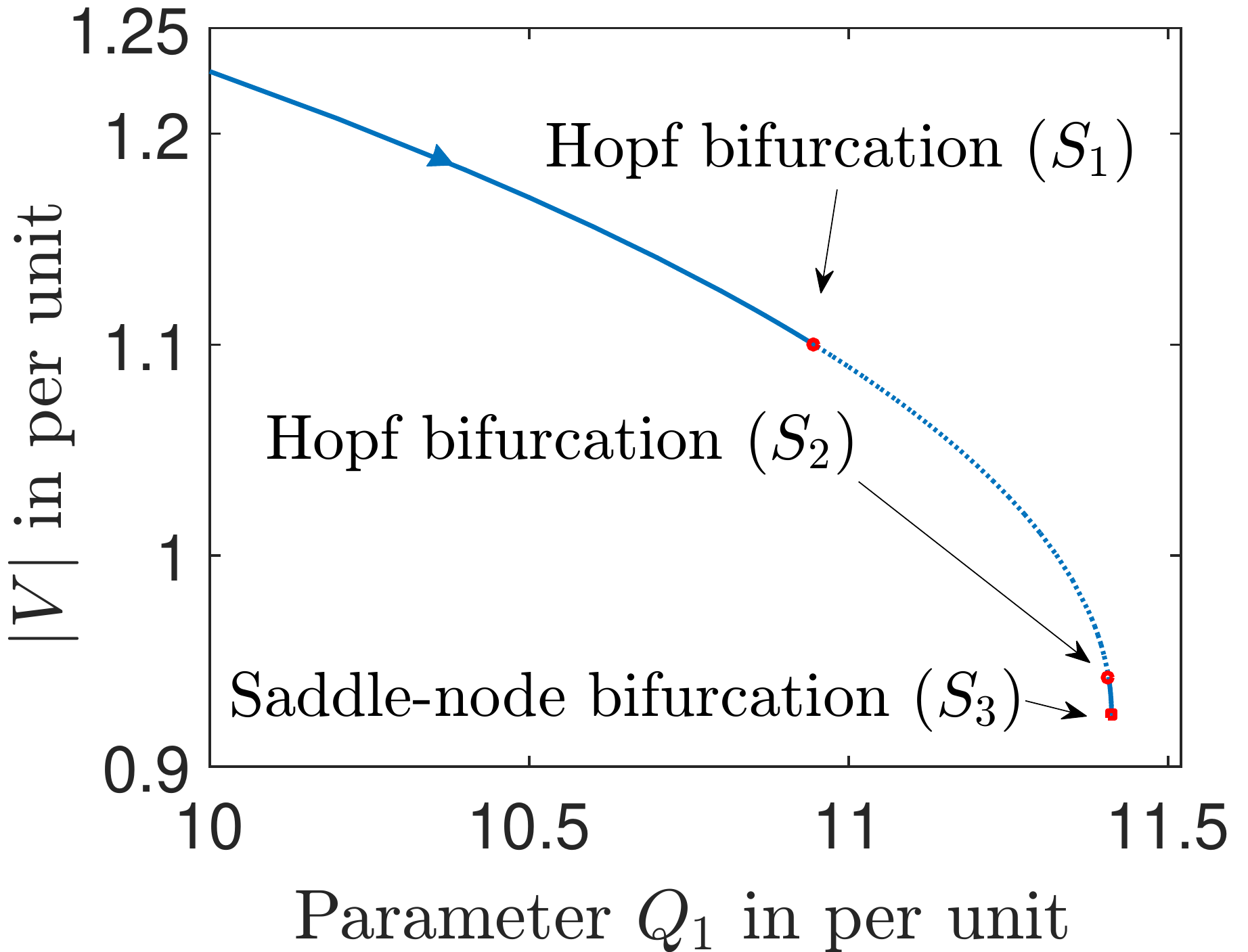}}
\caption{(a) 3-bus system. (b) Critical points of the system}\label{3_bus_fig}
\end{figure}

Consider a power network with two generators and a load, as shown in Fig. \ref{3_bus_fig}(a). The load is modelled as an induction motor in parallel with a constant $PQ$ load. The system is modelled as a four dimensional dynamical system with the state being generator angle $(\delta_g)$, generator angular velocity $(\omega)$, the load angle $(\delta_l)$ and magnitude of load voltage $(V)$. 
The dynamic equations for the system are
\begin{eqnarray}\label{eq_del}
&& \dot{\delta_g} = \omega\\ \nonumber
&& \dot{\omega} = 16.66667\sin (\delta_l - \delta_g + 0.08727)V\\ \label{eq_omega}
&& \qquad - 0.16667 \omega + 1.88074\\ \nonumber
&& \dot{\delta_l} = 496.87181 V^2 - 166.66667\cos (\delta_l - \delta_g \\ \nonumber
&& \qquad - 0.08727)V - 666.66667\cos (\delta_l - 0.20944)V\\ \label{eq_delL}
&& \qquad 93.33333V + 33.33333Q_1 + 43.33333\\ \nonumber
&& \dot{V} = -78.76384V^2 + 26.21722\cos (\delta_l - \delta_g \\ \nonumber
&& \qquad - 0.01241)V + 104.86887\cos (\delta_l - 0.13458)V\\ \label{eq_VL}
&& \qquad + 14.52288V - 5.22876Q_1 - 7.03268
\end{eqnarray}
For detailed analysis of the system equations we refer the interested reader to \cite{dobson_model, ajjarapu_bifurcation}.

The above power network has three \emph{critical points}, namely $S_1$, $S_2$ and $S_3$, as shown in Fig. \ref{3_bus_fig}(b). At $S_1$ and $S_2$, a pair of imaginary eigenvalues cross the imaginary axis and at $S_3$, a real eigenvalue becomes zero. Hence, the system becomes unstable at $S_1$, remains unstable from $S_1$ to $S_2$, then regains stability after $S_2$ and again becomes unstable at $S_3$. It is known that the instability at $S_1$ is angle instability and the instability at $S_3$ is voltage instability. In this section, we demonstrate how information transfer not only identifies the instability but also identify the states responsible for instability.

\subsection{State to State Information and Stability : 3-Bus System}

It is known that in the 3 bus network, there are both angle and voltage instability. In particular, the instability that occurs at $S_1$ (Fig. \ref{3_bus_fig}) is angle instability and the instability that occurs at $S_3$ is voltage instability \cite{ajjarapu_bifurcation}. 

\begin{figure}[htp!]
\centering
\subfigure[]{\includegraphics[scale=.41]{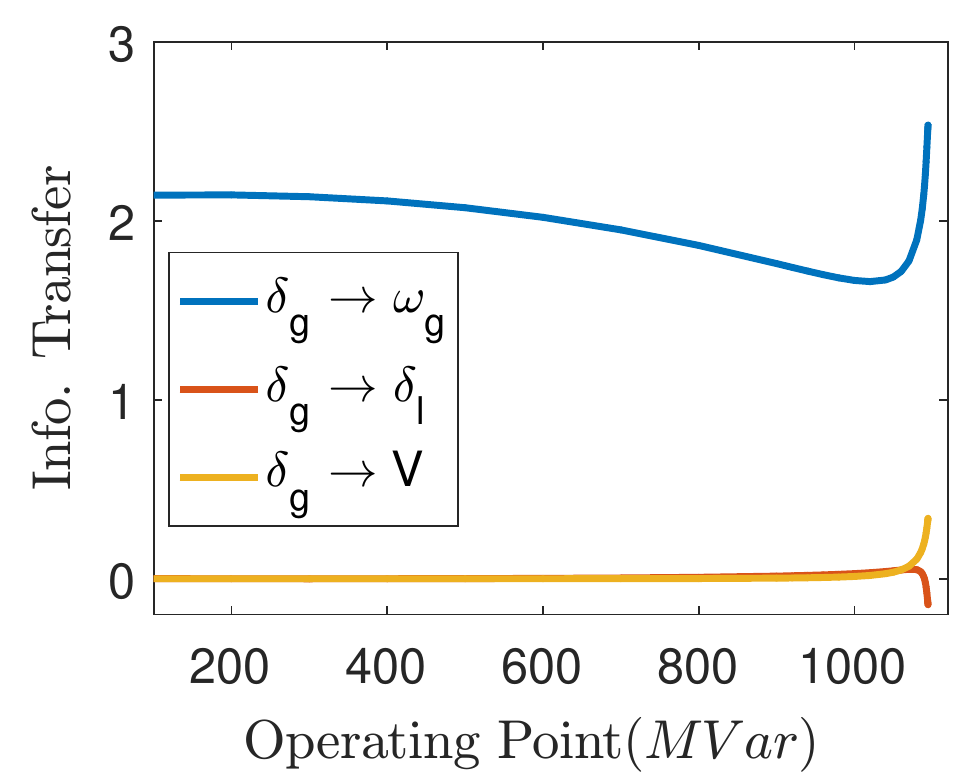}}
\subfigure[]{\includegraphics[scale=.42]{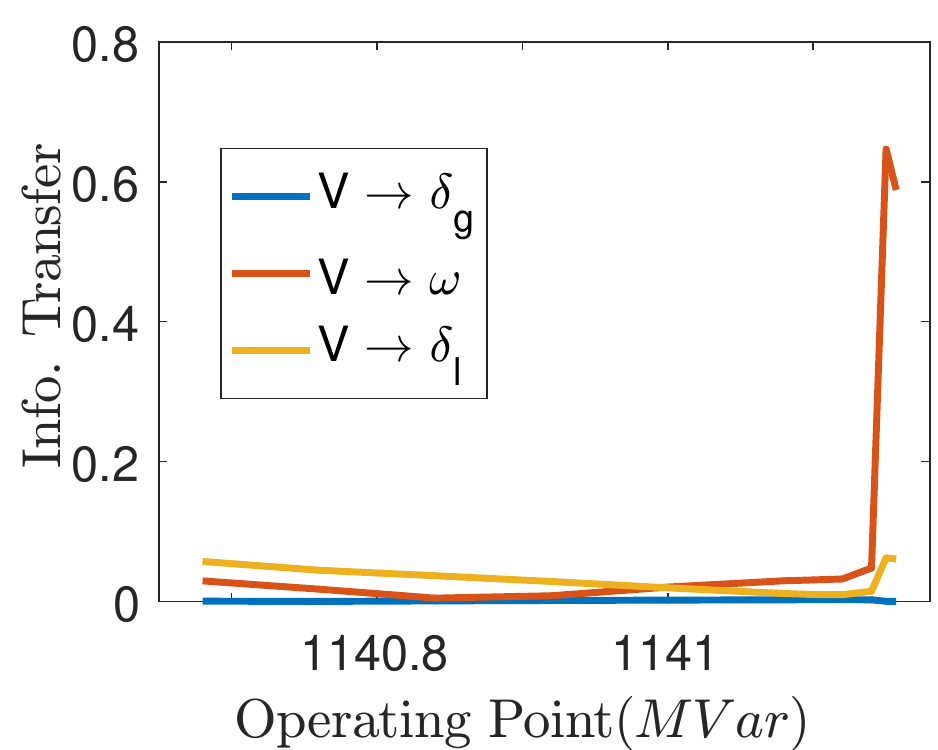}}
\caption{(a) IT from $\delta_g$ before $S_1$. (b) IT from $V$ before $S_3$.}\label{IT_3_bus}
\end{figure}  

In Fig. \ref{IT_3_bus}(a) we show the information transfer from the angle variable of the generator to all the other states, as the system approaches the first Hopf bifurcation. It can be seen that the information transfer from the angle variable shows a sudden increase. This is consistent with the fact that at $S_1$, the system undergoes voltage instability and it is the angle variable that causes the instability. Similarly, as the system approaches voltage instability $(S_3)$, the information transfer from the angle variable $(V)$ to all the other states shows the sudden increase. This is shown in Fig. \ref{IT_3_bus}(b). Hence, one can identify clearly the states which are responsible for the instability of the power network.

\subsection{Information Transfer \& Participation Factor: 3-Bus System}

For calculating the participation factors of each state of the $3$-bus system (\ref{eq_del})-(\ref{eq_VL}), we look at the linearized dynamics at the operating point $Q = 1141.1$ MVar. The rationale behind choosing this operating point is the fact that if $Q$ is increased further, the system undergoes Saddle Node Bifurcation (voltage instability). It is well-known \cite{ajjarapu_bifurcation} that at SNB, it is the load voltage $V$ that participates the most in the most unstable mode and this is shown in Table \ref{part_fact_unstable}.

{\small
\begin{table}[htp!]
\centering
\caption{Participation and Information Transfer to most unstable mode}\label{part_fact_unstable}
\begin{tabular}{|c|c|c|c|c|}
\hline
State.\textbackslash  Index & Participation Factor & Information Transfer  \\
\hline
    $\delta_g$ & $0.0137785$ & $0.04$ \\
     $\omega$ & $0.00046$ & $0.0094$\\
     $\delta_l$ & $0.00072$ & $0.0189$\\
     $V$ & $0.9850$ & $1.39884$\\
\hline
\end{tabular}
\end{table}}

Information transfer from the states to the most unstable mode, at this operating point, is shown in table \ref{part_fact_unstable}. For calculating the information transfer we discretized the linearized continuous-time model with $\Delta t = 0.001$ and calculated the steady state information transfer from the states to the most unstable mode. It is observed that both information transfer and participation factor identify the load voltage as the state responsible for instability. Not only does information transfer identify the load voltage as the state most responsible for instability, but it can be seen that the order of participation/influence is the same with both participation factor and information transfer. 
So far, we saw that participation factor and information transfer are doing the same thing. But one of the key advantages of the information transfer measure is the fact that one can look at information transfer between any combination of states (subspace) or from any combination of state to any output. Hence, in a system with a large number of states, one can treat each generator as a separate subspace and identify which generator is responsible for instability. Once the most critical generator is identified, one can zoom in to the states of the critical generator and identify the state(s) responsible for instability. This reduces the computation cost considerably. For more details on the relationship between participation factor and information transfer, we refer the interested reader to \cite{sinha_cdc_2017_power}.

\begin{algorithm}
\textbf{Given:} Power system dynamic model as \\
$\dot{x} = f(x,y)$ \hspace{0.2in}
$g(x,y) = 0$  \\
$ x \in \mathbb{R}^n$ are the dynamic states and $y \in \mathbb{R}^l$ are the algebraic states of the system.  \\
\textbf{Creating system matrices}
\begin{enumerate}
\item At a given operating point $(x_o, y_o)$, linearize the system by computing $\{ \bar A, \bar B, \bar C, \bar D \}$ as:\\
$\bar A = \frac{\delta f(x,y)_{(x_o, y_o)}}{\delta x} $  \hspace{0.2in}
$\bar B = \frac{\delta f(x,y)_{(x_o, y_o)}}{\delta y} $ \\
$\bar C = \frac{\delta g(x,y)_{(x_o, y_o)}}{\delta x} $  \hspace{0.2in}
$\bar D = \frac{\delta g(x,y)_{(x_o, y_o)}}{\delta y} $ 

\item Reducing system matrix
\begin{equation*}
\dot x = A_{cont}x = \{ \bar A - \bar B.(\bar D^{-1} \bar C) \} x.
\end{equation*}
\item Obtain an equivalent discrete dynamic model, using time step $\tau$ as:
\begin{equation*}
A = A_{disc} = e^{A_{cont} \tau}.
\end{equation*}

\hspace*{-0.4 in} \textbf{Computing information transfer}
\item To compute $T_{x_1 \rightarrow y}$, rearrange $A$ as:
\begin{eqnarray*}
A = \begin{pmatrix}A_x&A_{xy}\\ A_{yx}&A_{y}\end{pmatrix}=\begin{pmatrix}A_{x_1}&A_{x_1x_2}& A_{x_1 y}\\A_{x_2x_1}&A_{x_2}& A_{x_2 y}\\ A_{y x_1}&A_{y x_2}& A_{y}\end{pmatrix}
\end{eqnarray*}
\hspace*{-0.4 in} 
\item Choose a positive constant $\sigma$ and solve the Lyapunov equation $\Sigma = A \Sigma A^\top+\sigma^2 I$, 
to compute the steady state covariance matrix $\Sigma$.
\item Compute steady state information transfer $T_{x_1 \rightarrow y}$ as:
\begin{eqnarray*}
[T_{x_1\to y}]_t^{t+1}=\frac{1}{2}\log \frac{|A_{yx}\Sigma^s_yA_{yx}^\top +\sigma^2 I |}{|A_{yx_2}(\Sigma_y^{s})_{yx_2}A_{yx_2}^\top+\sigma^2 I|}
\end{eqnarray*}
where $\Sigma^s_y$ and $(\Sigma_y^{s})_{yx_2}$ are defined as in theorem \ref{IT_theorem}.
\end{enumerate}

\caption{Information transfer computation in power system}
\label{algo_IT}
\end{algorithm}

\section{IEEE 39 Bus System}\label{section_39_bus}
The model used in this section is based on the modeling described in \cite{Sauer_pai_book}. The power network is described by a set of differential algebraic equations (DAE) and the power system dynamics is divided into three parts: differential equation model describing the generator and load dynamics, algebraic equations at the stator of the generator and algebraic equations describing the network power flow. We consider a power system model with $n_g$ generator buses and $n_l$ load buses. The generator dynamics at each generator bus can be represented as a $4^{th}$ order dynamical model:

{\small
\begin{eqnarray*}\label{generator_dynamic_eq}
\begin{aligned}
&\frac{d\delta_i}{dt}  = \omega_i - \omega_s \\
&\frac{d\omega_i}{dt}  = \frac{T_{m_i}}{M_i} - \frac{E_{q_i}^{\prime} I_{q_i}}{M_i} - \frac{(X_{q_i} - X_{d_i}^{\prime})}{M_i} I_{d_i} I_{q_i} - \frac{D_i (\omega_i-\omega_s)}{M_i} \\
&\frac{d E_{q_i}^{\prime}}{dt}  = -\frac{E_{q_i}^{\prime}}{T_{{do}_i}^{\prime}} - \frac{(X_{d_i} - X_{d_i}^{\prime})}{T_{{do}_i}^{\prime}} I_{d_i} + \frac{E_{{fd}_i}}{T_{{do}_i}^{\prime}} \\
&\frac{dE_{{fd}_i}}{dt}  = -\frac{E_{{fd}_i}}{T_{A_i}} + \frac{K_{A_i}}{T_{A_i}} (V_{{ref}_i} - V_i) 
\end{aligned}
\end{eqnarray*}}
The algebraic equations at the stator of the generator are: 
{\small
\begin{align}\label{stator_algebraic_eq}
\begin{split}
& V_i \sin(\delta_i - \theta_i) + R_{s_i} I_{d_i} - X_{q_i} I_{q_i}  = 0 \\
& E_{q_i}^{\prime} - V_i \cos(\delta_i - \theta_i) - R_{s_i} I_{q_i} - X_{d_i}^{\prime} I_{d_i}  = 0 \\
& \qquad \text{for}\quad  i = 1,\dots, n_g.
\end{split}
\end{align}}
The network equations corresponding to the real and reactive power at generator and load buses are shown below. 
{\small
\begin{eqnarray}\label{network_eq}
\begin{aligned}
& I_{d_i} V_i \sin(\delta_i - \theta_i) + I_{q_i} V_i \cos(\delta_i - \theta_i) + P_{L_i} (V_i) \\ 
& - \sum_{k=1}^{\bar n} V_i V_k Y_{ik} \cos(\theta_i - \theta_k - \alpha_{ik}) = 0 \\ 
&  I_{d_i} V_i \cos(\delta_i - \theta_i) - I_{q_i} V_i \sin(\delta_i - \theta_i) + Q_{L_i} (V_i) \\ 
& - \sum_{k=1}^{\bar n} V_i V_k Y_{ik} \sin(\theta_i - \theta_k - \alpha_{ik}) = 0 \\
& \text{for} \;\ i = 1,\dots, n_g. \\
& P_{L_i}(V_i) - \sum_{k=1}^{\bar n} V_i V_k Y_{ik} \cos(\theta_i - \theta_k - \alpha_{ik}) = 0 \\
& Q_{L_i}(V_i) - \sum_{k=1}^{\bar n} V_i V_k Y_{ik} \sin(\theta_i - \theta_k - \alpha_{ik}) = 0 \\
& \text{for} \;\ i = n_g +1,\dots, n_g + n_l. 
\end{aligned}
\end{eqnarray}}
here, $\delta_i$, $\omega_i, E_{q_i}$, and $E_{{fd}_i}$ are the dynamic states of the generator and correspond to the generator rotor angle, the angular velocity of the rotor, the quadrature-axis induced emf and the emf of fast acting exciter connected to the generator respectively. The algebraic states $I_{d_i}$ and $I_{q_i}$ are the direct-axis and quadrature-axis currents induced in the generator respectively. Each bus voltage and its angle are denoted by $V_i$ and $\theta_i$. The parameters $T_{m_i}, V_{{ref}_i}, \omega_s, M_i$, and $D_i$ are the mechanical inputs and machine parameters applied to the generator shaft, reference voltage, rated synchronous speed, generator inertia, and internal damping. 
The stator internal resistance is denoted by $R_{s_i}$ and $X_{q_i}$, $X_{d_i}$, $X_{d_i}^{\prime}$ are the quadrature-axis salient reactance, direct-axis salient reactance and direct-axis transient reactance respectively. The exciter gain and time-constant are given by $K_{A_i}$ and $T_{A_i}$.

A power system stabilizer (PSS), that acts as a local controller to the generator is designed based on the linearized DAEs. The input to the PSS controller is $ \omega_i(t)$ and PSS output, $V_{{ref}_i}(t)$, is fed to the fast acting exciter of the generator. An IEEE Type-I PSS is considered here which consists of a wash-out filter and two phase-lead filters. The transfer function of PSS is given as follows

\begin{align}
\frac{\Delta V_{{ref}_i}(s)}{\Delta \omega_i(s)} = k_{pss} \frac{(1+sT_{num})^2}{(1+sT_{den})^2} \frac{s T_w}{1+sT_w} 
\label{eq_pss}
\end{align} 
where $k_{pss}$ is the PSS gain, $T_w$ is the time constant of wash-out filter and $T_{num}, T_{den}$ are time constants of phase-lead filter with $T_{num} > T_{den}$. 

Elimination of the algebraic variables by Kron reduction, generates a reduced order dynamic model given by $\Delta \dot{x}_g  = {A_{gg}} \Delta x_g + E_1 \Delta \tilde{u}$ where $\Delta {x}_g \in \mathbb{R}^{{7n_g}}$ and $\Delta \tilde{u} \in \mathbb{R}^{{n_g}}$. 

\begin{figure}[htp!]
\centering
\includegraphics[scale=.25]{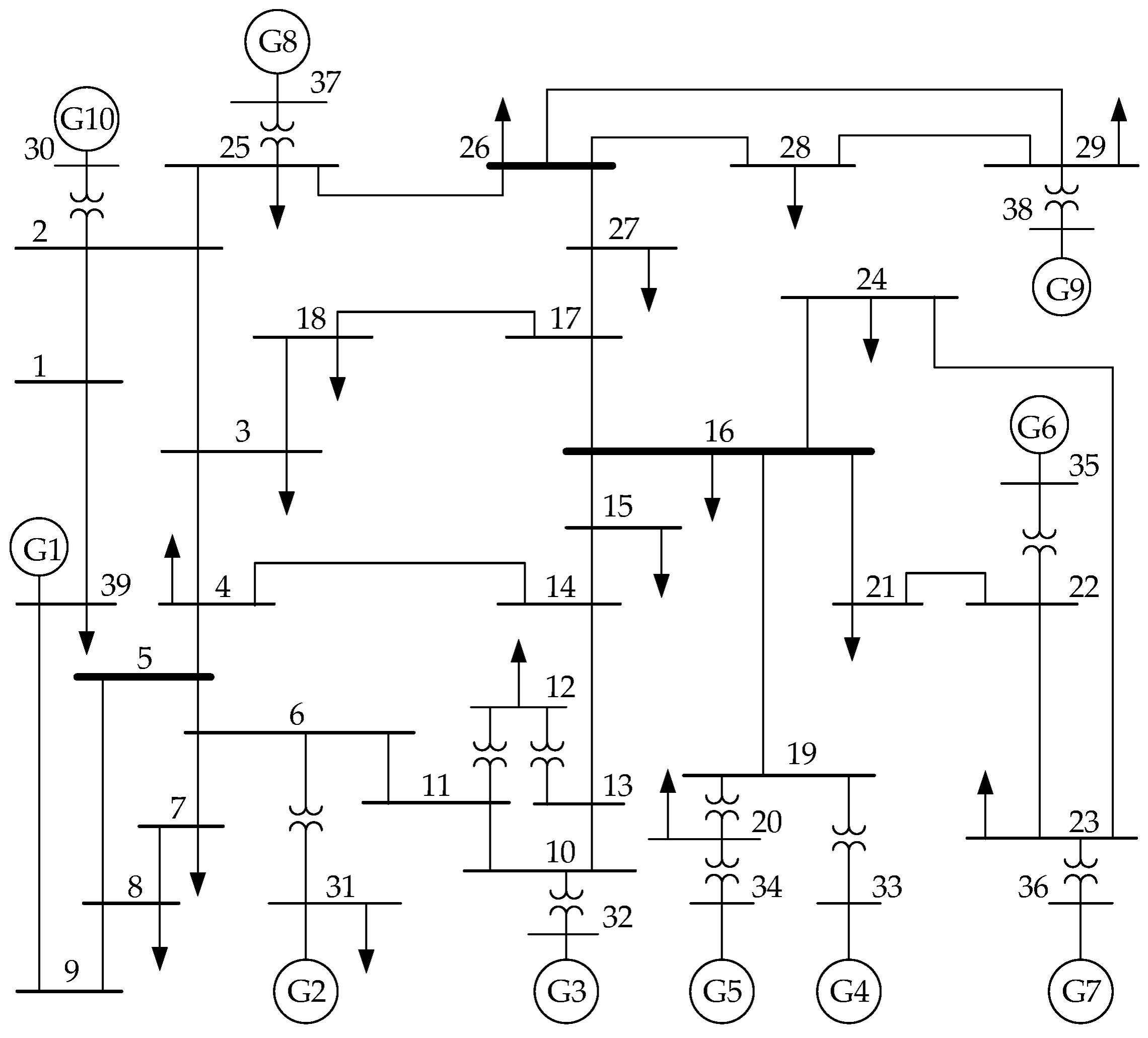}
\caption{IEEE 39 bus system.}\label{39_bus_fig}
\end{figure}

In this section, we consider the IEEE 39 bus system, the line diagram of which is shown in Fig. \ref{39_bus_fig}

The objective is to use the information transfer measure to identify the generator and the states of the generator which are responsible for instability. For identifying the states responsible for instability, the concept of participation factor is generally used. But one drawback of participation factor is the fact that it can not compute the participation of a combination of states to any particular mode. However, our formulation of information transfer is free from such issues and one can compute the information transfer from any combination of states to any other combination of states. As such, one can combine the states of each generator together and look at the information transfer between the generators. 

In the first set of simulations, we look at the information transfer between the generators and identify the generator from which the information transfer shows a sudden increase, as the network approaches instability. For simulation purposes, we linearized the system along the PV curve and used the linearized model to compute the information transfer between the generators. The discretization step used was 0.2 secs. 

\begin{figure}[htp!]
\centering
\subfigure[]{\includegraphics[scale=.43]{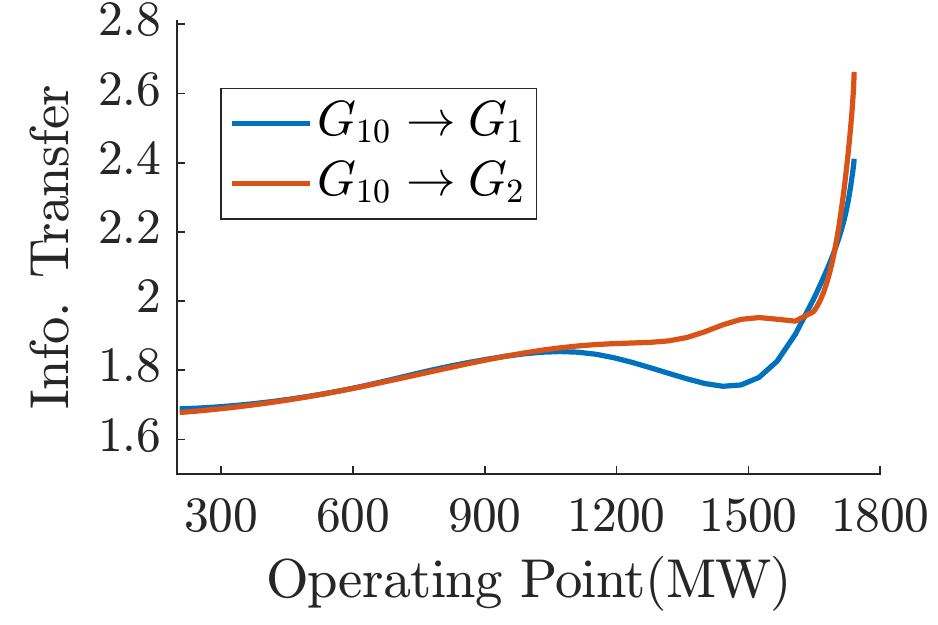}}
\subfigure[]{\includegraphics[scale=.4]{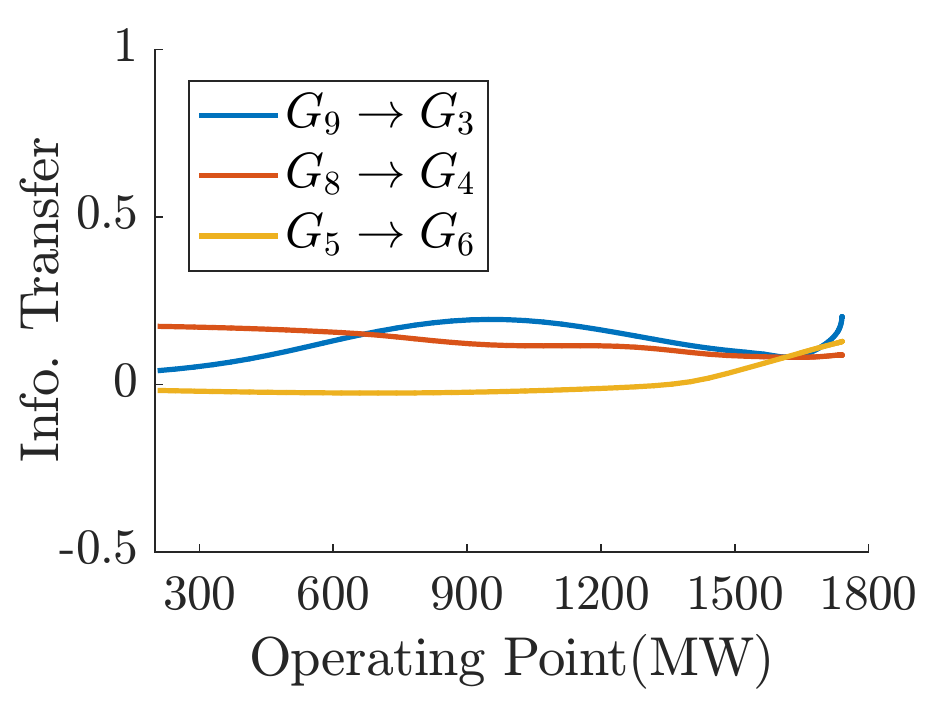}}
\caption{(a) Information transfer from generator 10 to generators 1 and 2. (b)Information transfer from generators 5, 8 and 9 to generators 3, 4 and 6 respectively.}\label{IT_from_G10}
\end{figure}

Simulation results identified generator 10 as the generator which is responsible for instability. This is because, the information transfer from generator 10 showed a sudden rise, as the network approached instability. Information transfer from generator 10 to generators 1 and 2 is shown in Fig. \ref{IT_from_G10}(a) and from the figure it can be clearly seen that the information transfer from generator 10 starts to increase rapidly. For comparison, we also plot the information transfer from generators 5, 8 and 9 to generators 3, 4 and 6 respectively in Fig. \ref{IT_from_G10}(b). It can be clearly seen in Fig. \ref{IT_from_G10}(b) that the information transfers between these generators do not change that much, thereby implying that these generators are not responsible for instability.

Once we have identified the generator which is most responsible for instability, we zoom in to the generator and identify the states of the generator which are responsible for instability. We have considered only the four states of the generator and not the states of the PSS. In particular, we have studied the information transfer from the states of generator 10 to other generators and identified the states of generator 10, the information transfer from which to other generators show the sudden increase.

\begin{figure}[htp!]
\centering
\subfigure[]{\includegraphics[scale=.38]{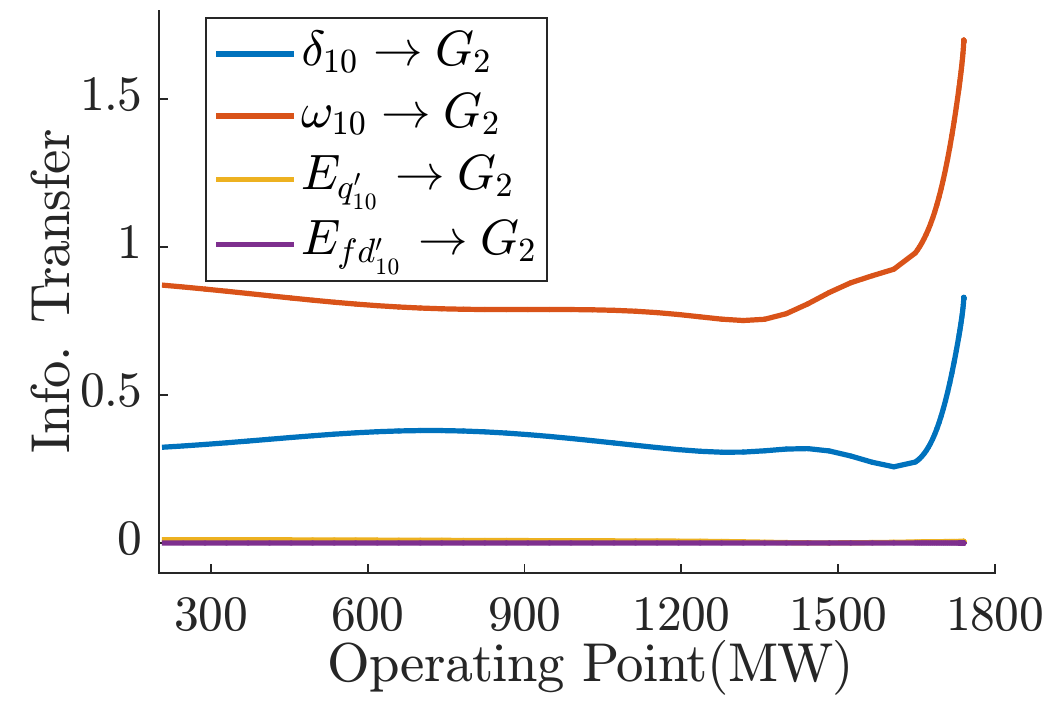}}
\subfigure[]{\includegraphics[scale=.4]{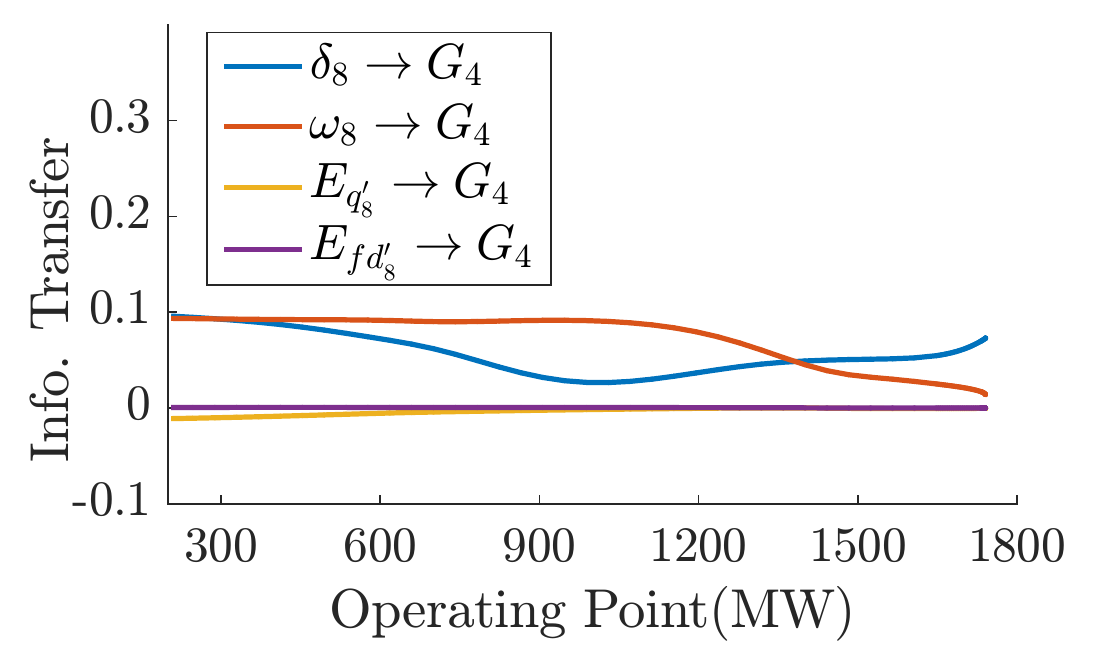}}
\caption{(a) Information transfer from states of generator 10 to generators 2. (b) Information transfer from states of generators 8 to generator 4.}\label{IT_G_10_zoomed}
\end{figure}

 In Fig. \ref{IT_G_10_zoomed}(a) we plot the information transfer from the individual states of generator 10 to generator 2. From the figure, it can be seen that the information transfer from the angle and the angular speed variables show a sudden increase, whereas the other transfers remain almost same throughout the PV curve. Hence, we conclude that it is the angle and the angular speed variables which are most responsible for instability. This fact is also verified by participation factor analysis, where it was found that the participation factor of the angle and angular speed of generator 10 to the most unstable mode is significantly high. For comparison, we also plot the information transfer from the states of generator 8 to generator 4 in Fig. \ref{IT_G_10_zoomed}(b). As expected, we find that the information transfer does not change much throughout the PV curve, and hence reaffirming the fact that generator 8 is not responsible for instability. The same conclusion holds for the states of the other generators as well.
 
One of the key developments in the power industry in recent times is the synchrophasor technology. Phasor measurement units (PMUs) are capable of sampling system variables down to the level of hundredth of a second. Hence, there is a strong need for a framework to articulate system stability from the available measurements. In \cite{sinha_acc_data,sinha_data_arxiv}, we have provided a data-driven approach to compute the information transfer measure from time series data. The data-driven approach for stability inference and characterization of influence in power networks is left for future investigations.

\section{Conclusion}\label{section_conclusion}
In this paper, we introduce a new measure of information transfer in a dynamical system for stability characterization in power network. In particular, we provide a conceptual basis on how causality characterization between dynamic states can reveal information about system instability and participating entities. Further, we illustrated the theory on 3 bus power network and recovered the known results on the cause of instability, voltage or angle, in the system. We compare information transfer concept with the existing method of participation factor and show how information transfer measure captures non-zero influence when participation factor fails to do so. Moreover, we illustrated the theory on IEEE 39 bus system and identified the states and the generators responsible for the instability of the 39 bus system. Future research efforts will focus on proposing a data-driven approach for computing information transfer in power network using results developed in \cite{sinha_acc_data,sinha_data_arxiv}.

\bibliographystyle{IEEEtran}
\bibliography{ref1,subhrajit_ref,subhrajit_power2}

\end{document}